# Layer dependence of graphene-diamene phase transition in epitaxial and exfoliated few-layer graphene using machine learning

Filippo Cellini[1§], Francesco Lavini[1,2§], Claire Berger[3,4], Walt de Heer[4,5], and Elisa Riedo[1*]

[1] Tandon School of Engineering, New York University, Brooklyn, NY, USA.

[2] Graduate School of Art and Sciences, New York University, New York, NY, USA.

[3] Institut Néel, CNRS-University Grenoble-Alpes, Grenoble, France

[4] School of Physics, Georgia Institute of Technology, Atlanta, GA, USA

[5] TICNN, Tianjin University, Tianjin, China

[*] Corresponding author: elisa.riedo@nyu.edu

[§] These Authors contributed equally (see Author contributions)

**Abstract.** The study of the nanomechanics of graphene - and other 2D materials - has led to the discovery of exciting new properties in 2D crystals, such as their remarkable in-plane stiffness and out of plane flexibility, as well as their unique frictional and wear properties at the nanoscale. Recently, nanomechanics of graphene has generated renovated interest for new findings on the pressure-induced chemical transformation of a few-layer thick epitaxial graphene into a new ultra-hard carbon phase, named diamene. In this work, by means of a machine learning technique, we provide a fast and efficient tool for identification of graphene domains (areas with a defined number of layers) in epitaxial and exfoliated films, by combining data from Atomic Force Microscopy (AFM) topography and friction force microscopy (FFM). Through the analysis of the number of graphene layers and detailed Å-indentation experiments, we demonstrate that the formation of ultra-stiff diamene is exclusively found in 1-layer plus buffer layer epitaxial graphene on silicon carbide (SiC) and that an ultra-stiff phase is not observed in neither thicker epitaxial graphene (2-layer or more) nor exfoliated graphene films of any thickness on silicon oxide ($SiO_2$).

# Introduction

Understanding the mechanical properties of two-dimensional (2D) materials is of fundamental importance for their application in development of new devices [1], especially in the presence of extreme requirements, such as for development of advanced flexible electronics [2] or aerospace systems [3]. The knowledge of the stiffness of 2D films, as well as their strength and hardness, can inform important decisions in materials selection, as fabrication technologies are shifting from small scale and small areas exfoliated flakes [4] to large area CVD and epitaxial films [5-7], and large volume batch productions and industrial applications are becoming more and more viable [8, 9].

Mechanical and nanomechanical testing of 2D materials is also a tool for materials discovery. Recent research efforts have focused on the pressure induced formation of new phases in 2D materials. In particular, the formation of an ultrastiff phase, diamene, was found while locally pressurizing epitaxial graphene films [10, 11]; other reports have shown the pressure induced formation of an insulating phase from exfoliated graphene flakes in a humid environment [12, 13], and more recently the formation of bonitrol from hexagonal boron nitride was demonstrated [14]. These films present structural, electrical, and mechanical properties completely different from their original 2D counterparts, and their stabilization in the form of two-dimensional layered structure can open avenues to new groundbreaking developments in nanotechnology. However, while important discoveries and new potential applications have been made possible by the accurate testing and understanding of the mechanics and force-induced structural modifications of 2D films, and specifically graphene, a framework to clearly correlate mechanical behavior with morphological properties, such as number of layers and type of graphene-substrate interface, still needs to be established.

Machine learning techniques offer new opportunities to gain insight in experimental data [15] and have been leveraged in the study of nanomaterials, including graphene. For example, deep learning techniques are gaining traction in scanning transmission electron microscopy (STEM) to identify and classify defects and vacancies in atomically thin crystals, and recent studies have focused on their application in 2D materials and graphene [16]. Unsupervised learning methods can also play an important role in materials research, as they allow estimation of structures in data without previous knowledge on the dataset [17, 18]. Cluster analysis is probably one of the most established areas of unsupervised learning, whereby subgroups or clusters are defined in the data by leveraging the pairwise similarities among their elements. Clustering techniques, such as K-means, have already been employed to identify graphene domains in Raman mapping [19], while the non-negative matrix factorization algorithm (NMF) has been used to decompose the Raman spectrum of epitaxial graphene in the pure graphene and silicon carbide (SiC) spectra [20].

In this paper, by means of atomic force microscopy (AFM) we correlate the number of layers in epitaxial graphene on silicon carbide (SiC) and exfoliated graphene on silicon oxide (SiO$_2$) with the frictional properties, transverse elasticity, and formation of ultra-stiff diamene layers. To this aim, we employ a fast and robust clustering technique, namely, spectral clustering [21], to identify the number of layers in the different domains in continuous epitaxial graphene films, and their spatial distribution based on nanomechanical measurements. By combining this machine learning driven AFM mapping with Å-indentation measurements [22, 23], we univocally detect domains/regions with a given number of layers, and relate this layer-number to the transverse elasticity of the specific domains/regions, opening avenues to machine-based (automatic) mapping and classification of graphene-diamene domains, even in the case of epitaxial graphene with complex morphologies. Through this analysis, we univocally demonstrate that the formation of ultra-stiff diamene is obtained only in 1-layer plus buffer layer epitaxial graphene on SiC and that thicker epitaxial graphene (2-layer plus buffer or more) and exfoliated graphene films on SiO$_2$ for any number of layers do not exhibit the formation of this ultra-stiff phase.

## Results

**Optical microscopy and Raman spectroscopy of exfoliated and epitaxial graphene**

Exfoliated and epitaxial graphene samples are prepared following the procedures described in the Methods section. Optical micrographs and Raman spectra are used to identify regions with different number of layers in exfoliated graphene flakes on SiO$_2$, as well as on continuous epitaxial graphene films grown on SiC. Figure 1(a) displays the optical micrograph of a 1-layer graphene flake on silicon oxide (285 nm oxide thickness). The large area of the flake allows for simple identification of the Raman spectra (532 nm laser). The spectrum of the 1-layer graphene is displayed in Figure 1(d). Following previous reports [24, 25], we are able to identify two characteristic peaks of 1-layer exfoliated graphene, namely the G peak located at ~1600 cm$^{-1}$ and the 2D peak located at ~2700 cm$^{-1}$ (2D/G integral intensity ratio is 2.8). According to literature [24, 25], the proportion between the intensities of the two peaks is usually above 2, with the intensity of the 2D peak being stronger than the G peak. This proportion is not observed in graphene flakes with more than one layer.

Figure 1(b) displays the optical micrograph of a multilayer flake. Several 2D structures can be observed at the same time in the micrograph according to the different gradation of color, namely 1-, 2-, 3- layer, as well as multilayer graphene [26, 27]. While the different structures can be observed in the optical micrograph, the definitive assignment of the number of layers can be performed only analyzing the Raman spectra of the different regions. Figure 1(e) displays the Raman spectra collected in different regions of the flake, which are marked in Figure 1(b) following the same color coding. Notably, a large internal area of the flake displays spectra characteristic of 1-layer graphene (green marker, 2D/G integral intensity ratio is 2.9), while neighboring regions present spectra characteristic of 2-, 3-layer graphene (red marker, 2D/G integrated intensity ratio is 2.3) and multi-layer graphene (black marker, 2D/G integrated intensity ratio is 1.6).

Figure 1(c) displays the typical optical micrograph of a continuous epitaxial graphene film surface, where several continuous lines can be identified at approximately 45° orientation corresponding to the steps generated during high temperature annealing on silicon carbide. The Raman spectrum of epitaxial graphene is displayed in Figure 1(f). In the case of epitaxial graphene grown on silicon carbide, the emission spectrum of the substrate partially overlaps with the emission of the graphene layers [28]. Therefore, the spectrum of epitaxial graphene in Figure 1(f) shows several peaks in the range 1500-1800 cm$^{-1}$ that are associated with the Raman signature of silicon carbide [29]. In order to visually highlight and isolate the epitaxial graphene characteristic G and 2D peaks, the spectrum of bare silicon carbide annealed in Argon is also reported in Figure 1(f). The characteristic peaks of graphene are identified in the spectrum in Figure 1(f) at ~1600 cm$^{-1}$ (G peak) and ~2720 cm$^{-1}$ (2D peak). Notably, the ratio between the 2D and G peaks is approximately 0.8 (after subtraction of the silicon carbide spectrum), which is far from the ratio measured for 1-layer exfoliated graphene (the 2D/G integral intensity ratio is 1.92 for the epitaxial graphene film). The 2D peak is also substantially broader than for 1-layer exfoliated graphene. The difference between the spectra of epitaxial graphene and exfoliated layers can be explained by considerations on the nature of these films: i) graphene flakes can be spatially isolated to a size that is easily probed by traditional Raman microscopes (~1 μm); on the other hand, epitaxial graphene on silicon carbide presents a complex structure, where domains of buffer layer (BfL), 1-layer, and 2-layer graphene coexist inside regions with size of few micron squared; ii) a strong interaction exists between the substrate (silicon carbide) and the first carbon layer (buffer layer, BfL), which may result in a broadening and shifting of the peak due to the presence of residual in plane strains in this structure.

The values of the full width at half maximum (FWHM) are computed for the G and 2D peaks for all the spectra collected for exfoliated and epitaxial graphene samples. For single-layer graphene in Figure 1(a), the value of FWHM is 12.6 cm$^{-1}$ for the G peak and 30.8 cm$^{-1}$ for the 2D peak. For single-layer exfoliated graphene in Figure 1(b), the value of the FWHM is 10.1 cm$^{-1}$ for the G peak and 29.0 cm$^{-1}$ for the 2D peak; for 2-3 layer graphene is Figure 1(b), the value is 12.2 cm$^{-1}$ for the G peak and 57.4 cm$^{-1}$ for the 2D peak; for multi-layer graphene is Figure 1(b), the FWHM is 16.2 cm$^{-1}$ for the G peak and 77.0 cm$^{-1}$ for 2D peak. These values are directly computed from the experimental curves without deconvoluting the multiple Lorentzian components associated to the different Raman modes observed in graphene with more than one layer [24]. Values of FWHM for exfoliated graphene flakes are comparable with previous reports [24], where the resulting FWHM computed for the 2D peak of single-layer graphene is close to 25 cm$^{-1}$, while the width of the 2D peak progressively broaden with increasing thickness of the 2D layers due to the contribution of additional vibrational modes to the Raman scattering of graphene with more than one layer. From Figure 1(c), the values of FWHMs for the epitaxial graphene films are 15.6 cm$^{-1}$ for the G peak and 51.0 cm$^{-1}$ for the 2D peak, which are comparable with spectral data reported in the literature for few-layer graphene films on SiC [30, 31]. The spectrum of epitaxial graphene is obtained from the contributions of a mixture of 1-layer and 2-layer graphene films that are distributed over the SiC surface in regions that are smaller than the laser spot. For this reason, the isolation of 1-layer and 2-layer domains in epitaxial graphene films is a challenging task using the sole information provided through Raman microscopy.

**AFM morphology and friction map of exfoliated graphene on SiO$_2$**

Topography of few-layer exfoliated graphene flakes on SiO$_2$ presents large plateaus corresponding to the 2D layers [32, 33]. The topography measured in contact mode for 1-2-3 graphene layers from the flake in Figure 1(b) is displayed in Figure 2(a) and (c). The steps separating the different layers are sharp and are easily identifiable in the topography. The height of the steps is in the range 0.4-0.5 nm, which is in line with the expected thickness of the graphene layers, that is, 0.34 nm.

Frictional forces measured in contact mode are inputted together with topography data to segment the surface using the spectral clustering algorithm, see Methods section. Figure 2(b) displays the frictional force map together with the results of the clustering, where clustered regions are superimposed to the FFM scan in what we call the spectral clustering (SClust) frictional map. As expected, results of the clustering identify three well separated regions corresponding to the three graphene layers. Data from the clustering for both height and frictional (lateral) force are displayed in Figure 2(d). Notably, the region corresponding to 1-layer graphene presents an average friction force of 1.6 nN with an associated standard deviation of 0.3 nN, while the relative height of the region (assumed as the reference plane) is 0.0 nm with standard deviation 0.1 nm. The region corresponding to the 2-layer graphene has average frictional force of 0.9±0.2 nN and average height of 0.3±0.1 nm. The region of 3-layer graphene has average frictional force of 0.6±0.2 nN and average height of 0.9±0.1 nm.

These results are in line with previous observations [33-35], whereby a progressive decrease of the frictional force is observed in exfoliated graphene with increasing number of layers. In particular, Reference [34] shows a quasi-exponential decrease of the frictional force with the number of layers. This behavior is explained in Reference [33] through the so-called puckering effect. In this model, the single layer graphene, loosely attached to the underlying silicon oxide substrate, would bend under the pressure applied by the AFM tip creating an increased contact region that, in turn, would result in a higher lateral (frictional) force transmitted to the AFM tip. By increasing the thickness of the graphene sheet, bending of the 2D layer is reduced and consequently puckering of the layers, which explains the reduction of the frictional force with increasing number of layers, as we observe in our experiments.

**AFM morphology and friction map of epitaxial graphene on SiC**

The morphology of graphene epitaxially grown on SiC is more complex than the simple layered structure of exfoliated graphene flakes transferred on a wafer substrate. In this work, we investigate the properties of state-of-the-art epitaxial graphene samples fabricated following the original CCS methodology discussed in References [5, 36]. The complexity of the physical structure of these epitaxial graphene films derives from the sublimation process, which is a complex thermodynamic cycle controlled through several process parameters, whose contribution to the final product is still under investigation [5, 37]. The topography recorded during a set of experiments conducted on epitaxial graphene samples is displayed in Figure 3(a) and (c). From the image, we can observe few regions of higher elevation, which are locations of the surface where Si sublimation has not occurred or has occurred at substantial slower rates. These regions are usually identified as buffer layer (BfL), which is a first carbon layer with crystalline structure similar to the structure of graphene that is partially bonded to the underlying surface of SiC [38]. These regions are surrounded by regions of lower topography that are associated to the formation of 1-layer and, in less extended domains, 2-layer graphene on the Si face of SiC. In particular, 2-layer graphene domains are found in the more depressed areas of the surface, where sublimation of Si has been faster during processing. The complex 3D geometry of these layers together with the small variation in frictional forces observed between 1-layer and 2-layer graphene [39] may hinder the clear identification of the different graphene regions, calling for better tools for detection of the different domains.

To obtain a fast and robust identification of the 1-layer and 2-layer regions, we analyze frictional data through the spectral clustering algorithm. Results of clustering performed on the sole friction force data are displayed in Figure 3(b), and the frictional map is displayed together with the SClust map identified by the algorithm. Notably, the SClust frictional map gives a clear separation of the 1-layer and 2-layer regions, which is not readily observed in the original scan data. In addition, the algorithm provides a direct quantitative evaluation of frictional forces in the different regions, as shown in Figure 3(d). The region corresponding to the BfL presents an average friction force of 16.2 nN with an associated standard deviation of 1.9 nN. The region corresponding to 1-layer graphene on SiC presents instead an average frictional force of 1.7±0.5 nN. The third region (2-layer graphene on SiC) has average frictional force of 1.2±0.3 nN. Notably, the friction on the BfL is more than 10 times higher than friction of 1-layer graphene [39, 40]; the higher friction forces on BfL can be attribute to the roughness of this interface and the strong interaction with the SiC substrate [41]. In good accordance to the literature, a decrease of the frictional force is observed between 1-layer and 2-layer graphene. In experiments conducted in ultra-high vacuum [39], this variation has been attributed to a reduction of frictional force in 2-layer epitaxial graphene due to a change in phonon-electron coupling, which may be related to the progressive reduction of the interaction with the SiC substrate with increasing number of layers. However, given that experiments presented here are conducted in air, at humidity level of ~40% RH, the effect on friction forces of the tip-sample adhesion due to capillary interaction cannot be completely ruled out [42], as it may play an important role in determining the friction forces measured on graphene with different number of layers [43-47].

In epitaxial graphene, the formation of 1-layer, 2-layer, and BfL structures can occur independently on different SiC terraces during thermal sublimation of Si creating a quite complex layered architecture. By applying the clustered regions identified from friction map to topography data, a quantitative analysis of the height distribution is performed in a confined region within one of the terraces. We found that the average height of 1-layer graphene is -1.2±0.1 nm, the average height of the 2-layer region is -1.6±0.2 nm. The average height of BfL regions, assumed as the reference plane, is 0.0±0.2 nm. Notably, we obtain that the vertical distance from the exposed buffer layer to the nearby 1-layer region is approximately 1.2±0.3 nm, a value which accounts for the roughness of the buffer layer ($\approx$ 0.3 nm) [41, 48], the thickness of the newly formed graphene layer (1-layer, 0.34 nm) on top of BfL, and the thickness of the removed unit cell of 4H-SiC during sublimation ($\approx$ 1 nm) [49].

**Å-indentation of epitaxial graphene and exfoliated graphene vs. number of layers**

Å-indentation of graphene is performed to estimate the transverse elasticity of different graphene layers in both epitaxial and exfoliated films, as determined from the spectral clustering technique. Selective experiments are conducted to probe the stiffness of the different regions appearing in the SClust frictional map, namely BfL, 1-layer, 2-layer epitaxial graphene films and 1-,2-,3-layer exfoliated graphene flakes.

Å-indentation curves measured in the points marked on the SClust frictional map in Figure 4(a) for an epitaxial graphene film are displayed in Figure 4(b). Experimental results show that the number of graphene layers clearly controls the slope of the indentation curves in the different domains, with steeper indentation curves measured on 1-layer graphene (high stiffness) and softer indentation curves measured on 2-layer graphene (lower stiffness). The indentation curves of BfL range between indentation curves of 1-layer and 2-layer graphene. Stiffness of BfL is known to be comparable to the stiffness of freshly cleaved bare silicon carbide [10], since oxidation of external layers of SiC plays a role in reducing the mechanical stiffness of this substrate at the nanoscale after long exposition to air. Therefore, 1-layer epitaxial graphene exhibit stiffness at the nanoscale higher than the SiC substrate, as already discussed in our previous work [10, 11], while we report herein that the same behavior is not observed in the 2-layer graphene film. This ultra-stiffness may be attributed to the phase transition from 1-layer graphene on BfL to a new phase named diamene, whose structure and properties have been proposed in [10]. A direct comparison is conducted between the epitaxial graphene indentation curves and the indentation curves for other ultra-stiff materials, namely SiC ($E$~410 GPa), CVD diamond ($E$~1000 GPa), and Sapphire ($E$~400 GPa), also reported in Figure 4(b). Stiffness of 1-layer graphene measured in these experiments is substantially higher than the stiffness of sapphire and comparable or higher than the stiffness of diamond. Interestingly, 2-layer graphene is softer than sapphire and considerably softer than the SiC substrate/BfL, which shows how the formation of multilayer graphene progressively reduce the mechanical stiffness of the surface, with the transverse elasticity of 10-layers graphene being comparable to the stiffness of graphite [10, 22]. Nonlinear fitting of the indentation curves using the Hertz function gives an estimation of Young's modulus for the simplified Hertzian contact of 855±161 for 1-layer epitaxial graphene domains, 310±58 GPa for 2-layer epitaxial graphene, and 439±14 GPa for the BfL. The modulus is 453±61 GPa for sapphire (assumed as the reference material), 375±82 GPa for SiC, and 950±210 GPa for CVD diamond.

There are still many open questions regarding the stability, structure, and isolation of the pressure-induced diamene phase exhibiting mechanical response to indentation similar to that of diamond. In Figure 4(b), it is clear a substantial stiffening effect in the 1-layer + BfL graphene film on SiC as compared to both the pristine SiC substrate, and sapphire. Furthermore, 1-layer + BfL is consistently stiffer than the BfL and 2-layer graphene regions. However, the stiffness of the formed diamene layer is not uniform in all the positions tested, and there are points where 1-layer + BfL displays a stiffness larger than CVD diamond, and other points where it is smaller. This result may be related to the fact that the conversion from graphene to diamene may not be complete over the entire tip-sample contact area in all experiments [10, 11], and to experimental errors associated to the Å-indentation method. The structure of diamene phase is still under investigation, and different diamene structures may exhibit substantially different mechanical properties [10]. Additional work should be carried out in order to predict the properties of diamene and hetero-structures of diamene and graphene. Finally, it is worth noticing that the transverse elasticity of the film correlates with its frictional properties, whereby 1-layer epitaxial graphene on BfL presents both higher friction and higher stiffness than 2-layer epitaxial graphene structures.

Å-indentation curves of exfoliated graphene on $SiO_2$ are displayed in Figure 4(d). The locations on the flake where indentation experiments are performed are identified by the markers on the SClust frictional map in Figure 4(c). Data for indentation curves on bare $SiO_2$ ($E$~60 GPa) measured outside the flake region are also reported in Figure 4(d), together with indentation curves measured on bulk (0001) sapphire, which is used as an internal reference. Indentation data on exfoliated graphene on $SiO_2$ show that for all the layer-numbers investigated here (from 1- to 3-layer) the transverse stiffness of the exfoliated graphene/$SiO_2$ system is always lower or equal to that of bare $SiO_2$ substrate (i.e. without the flakes on top). The stiffness of the graphene films on silicon oxide is much lower than the stiffness of the sapphire substrate, due to the substantial difference in mechanical properties of sapphire ($E$~400 GPa) and silicon oxide ($E$~60 GPa). These results indicates a very different behavior compared to epitaxial graphene on SiC, where we observed for 1-layer plus BfL a very high stiffness, which is larger than the SiC substrate and even larger than CVD diamond films. Clearly, the graphene-diamene phase transformation is not occurring for the exfoliated flakes on $SiO_2$. The reason could be related to the different graphene-substrate interaction. We posit that the chemical and electronic structure of 1-layer epitaxial graphene sitting on top of the BfL, which is in part chemically bonded to the SiC substrate, promotes the formation of the ultra-hard diamene film through chemical modification of the graphene/BfL/substrate interfaces under pressure as observed in [10], and in [12] and [14] for graphene and other 2D films. The proposed mechanism for room-temperature formation of diamene from 1-layer plus BfL structures under pressure is a progressive re-hybridization of $sp^2$ 2D layers into $sp^3$ diamene, which is an ultra-stiff and ultra-hard structure. The formation of the diamond-like diamene is favored by saturation at the BfL/substrate interface of the dangling bonds formed during the pressure induced rearrangement of atoms in the 1-layer/BfL layers [10]. Saturation of dangling bonds is also possible through the formation of bonds with –H and –OH groups, which are naturally available at the interface of the 2D layers when experiments are conducted in air at moderate humidity levels (>RH 35%) [13, 14]. However, while it has been shown [13, 14] that formation of diamond-like structures alternative to diamene is possible through available –H/–OH contaminants during compression of exfoliated graphene layers on $SiO_2$, in our experiments at RH ~40% on exfoliated 1-layer graphene, we found that these structures do not display ultra-high stiffness nor hardness, which is line with what is generally observed for hydrogenated diamond-like carbon films [50].

Figure 5(a) and 5(c) displays the topographic scans of a 1-layer exfoliated graphene flake and few-layer exfoliated graphene flake showing regions of different thickness. The thickness of the single layer as well as the thickness of the multilayer regions is assessed from the topographic scan by direct comparison with the SiO$_2$ substrate, as reported in the cross-sectional topography displayed in Figure 5(a) and (c). Indentation curves measured in different points on the flakes are displayed in Figure 5(b) for the single layer and 5(d) for the multilayer flake. In addition, in Figure 5(d), we include data measured for a multilayer graphite flake (bulk>>10 layers). Qualitative analysis confirms the same result found in Figure 4(c)-(d), whereby stiffening of the surface is not observed in exfoliated graphene on silicon oxide. Based on our experiments, 1-layer curves show higher stiffness than 2-layer and >3-layer graphene, which suggests that increasing number of graphene layers progressively reduces the transverse stiffness of the film. By fitting the Hertz function on the indentation curves in Figure 5(d), we obtain an indentation modulus of 63±12 GPa for the SiO$_2$ substrate (which is assumed as the reference), while a modulus of 59±3 GPa, 46±8 GPa, and 39±6 GPa is obtained for 1-layer, 2-layer and >3-layer graphene films, respectively. The modulus of the bulk is estimated to be 33±5 GPa, which is close to the modulus expected for graphite ($E$~36 GPa). It is clear that what we observe is the superposition of two materials with different mechanical properties, multi-layer graphene (graphite) with transverse modulus of ~36 GPa and SiO$_2$ with modulus of ~60 GPa. Therefore, with increasing thickness of the top soft film (multi-layer graphene) the indentation modulus approaches the lower modulus of ~36 GPa.

## Conclusions

In this paper, through detailed Å-indentation experiments and machine learning clustering, we uncovered how the ultra-stiff diamene-graphene phase transition and interlayer elasticity depend on the graphene-substrate interaction and number of layers in epitaxial graphene grown on SiC and exfoliated graphene on SiO$_2$. The correlation of topography and friction force microscopy provides sufficient information to univocally identify the different graphene structures/number of layers in both exfoliated and epitaxial graphene. The use of spectral clustering techniques allows for fast and reproducible correlation of topographic and friction data and provide deeper insight in the analysis of graphene films, such as the identification of the number of layer and reconstruction of complex domains, even with limited friction contrast. Based on this methodology we conclude that the ultra-high stiffness can indeed be univocally attributed to 1-layer plus BfL epitaxial graphene, see Table 1. On the other hand, we also demonstrate that the stiffening effect is not observed in epitaxial graphene for a larger number of layers, precisely 2-layer plus BfL, or simple BfL or in exfoliated graphene on $SiO_2$ for any layer-number. We ascribe the lack of graphene-diamene phase transition in a larger number of layers for epitaxial graphene to the unlike possibility to tilt the planes in order to align for the A-A stacking [10]. On the other hand, the missing diamene formation when pressurizing exfoliated graphene on $SiO_2$ is associated to the effect of the substrate, which is unable to provide the electrons required to saturate the dangling bonds that are formed during re-hybridization of the $sp^2$ 2D layers into $sp^3$ diamene. While saturation of dangling bonds may be achieved through bonding with –H and –OH groups available at the 2D layers interface [13, 14], the resulting structures may not show similar stiffening effects as compared to diamene formation in 1-layer plus BfL epitaxial films.

|  | Epitaxial Graphene film | | | Exfoliated Graphene flake | |
|---|---|---|---|---|---|
|  | Modulus (GPa) | Friction (nN) |  | Modulus (GPa) | Friction (nN) |
| BfL | 439 ± 14 | 16.2 ± 1.9 | 1-layer | 59 ± 3 | 1.6 ± 0.3 |
| 1-layer | 855 ± 61 | 1.7 ± 0.5 | 2-layer | 46 ± 8 | 0.9 ± 0.2 |
| 2-layer | 310 ± 58 | 1.2 ± 0.3 | 3-layer | ~ 39 ± 6 | 0.6 ± 0.2 |

**Table 1.** Summary of the nanomechanical properties of epitaxial graphene films and exfoliated graphene flakes.
11

## Methods

**Growth of epitaxial graphene on SiC**

Continuous epitaxial graphene films are prepared by thermal sublimation of silicon from the surface of 4H-silicon carbide (SiC) wafers following the confinement-controlled sublimation (CCS) method described in References [5, 51, 52]. SiC wafers are cleaved into small samples of approximately 5 x 5 mm$^2$, which are polished on the growth face. The SiC samples are placed in a graphite crucible inside a high temperature IR furnace for graphene growth. Temperature and time are optimized to produce 1±1 layer of graphene on the SiC(0001) face (Si-face) [5]. Very importantly, we name "buffer layer" (BfL) the carbon interfacial layer between 1-layer graphene and the SiC substrate. To promote formation of high quality graphene films, the sublimation rate is controlled by using an Argon flow inside the crucible, while sublimated Si atoms are allowed to diffuse in the furnace chamber through a small ventilation hole. After CCS, the presence of few layer graphene structures on the Si-face of SiC is verified using Raman spectroscopy. Spectra of the surface are collected using a Horiba HR800 Raman microscope.

**Preparation of exfoliated graphene on SiO$_2$**

Graphene exfoliation is performed by adapting the process originally reported in Reference [53] and revised in References [54, 55]. Few-layer and single-layer graphene flakes are successfully transferred on a silicon oxide substrate (SiO$_2$ wafers, p-type, 285 nm oxide thickness, purchased from Graphene Supermarket) and identified using a Nikon Eclipse LV150N-CH optical microscope. Flakes of different thickness are identified using color variations due to optical interference, as described in References [26, 27]. The effective number of layers is verified in a separate set of experiments using Raman spectroscopy, as reported in [24, 25]. Spectra of the flakes are measured using a Horiba HR800 Raman microscope.

**AFM Friction force microscopy (FFM)**

Friction force microscopy (FFM) experiments are conducted on an Agilent Picoplus AFM using Nanosensors DT-NCHR polycrystalline diamond coated silicon tips are employed in all the experiments with resonant frequencies in the range 400-500 kHz, spring constant in the range 70-80 N/m, tip radius in the range 100-200 nm. Notably, while a stiff tip (high spring constant) does not guarantee optimal sensitivity in order to increase resolution in friction force microscopy, tips with a high spring constant are required for Å-indentation experiments and therefore are used in this work. The experiments are conducted in ambient conditions (RH ~40%, T ~25°C) in contact with normal force ranging between 20 and 300 nN and scan rates ranging between 0.5 and 1.5 Hz. Friction force data are acquired during the first few scans to reduce the effect of wear and tip contamination on the 2D graphene layers. Following References [56, 57], the frictional force is computed from experimental data by using the formula:

$$F_l = k_l Dx = \frac{wt^3 E}{6L^2 \left(\frac{t}{2} + h\right)} S_n V_l$$

where $k_l$ is the lateral spring constant of the cantilever, $Dx$ is the lateral torsion of the cantilever, $w$, $t$, and $L$ are the width, thickness and length and E is the Young's modulus of the AFM cantilever, respectively. The value of the AFM sensitivity $S_n$ is directly measured before the experiments by acquiring the force-distance curve, and it is in the range 50-60 nm/V. The voltage signal $V_l$ is the lateral deflection of the AFM cantilever that is measured as half the difference between the lateral deflections in the trace and retrace scans.

**Spectral clustering for morphological study of graphene**

A spectral clustering algorithm is employed to analyze the AFM data, detect the different graphene structures, and compute the mean values and distributions of the clustered regions [21, 58]. With respect to traditional methodologies, advanced clustering algorithms allow for fast, accurate, and robust determination of the material properties from AFM data in the presence of complex spatial distributions [59, 60]. Spectral clustering can be regarded as a generalization of the K-means algorithms to non-convex clusters [17], which are often found in complex datasets, such as those generated from epitaxial graphene samples whose surfaces are characterized by non-continuous variations in topography and nanomechanical properties. On the other hand, datasets from continuous exfoliated graphene samples tend to be convex, and separation of the clusters would be possible using K-means. Other approaches to non-convex clustering are for example diffusion learning methods [61, 62].

The spectral clustering algorithm that we developed is based on the Sklearn package in Python [63] for identification and clustering of the scan data. In a typical implementation, AFM data are fed to the Python script as a $4 \times N$ data vector (where N is related to the AFM image spatial resolution, N=65536 in our datasets), whose columns correspond to the x coordinate, y coordinate, topography (height), and friction/phase data of the scan, respectively. During preprocessing, the initial data set is reshaped in a $4 \times 256 \times 256$ data vector and a padding step with size 10 to 30 points is applied to remove possible corrupted data on the edges of the scan area. Flattening of the topography is usually necessary in our scans. Flattening is performed by subtracting the background, which is identified through fitting of the topography with a quadratic function in 2D (higher order functions may be required depending on the quality of the topography data). Notably, background subtraction is a necessary step to perform clustering in topography data where the AFM image presents substantial low frequency noise (tilt, bow,..). In some analysis, de-noising is also applied using a Gaussian filter with size ranging between 3 x 3 and 9 x 9, depending on the dataset. In addition, to reduce the size of the dataset, a moving window and/or a maxpool filter can be selected. The second approach is shown to be particularly effective to reduce the dimension of the friction data vector.

After preprocessing, each column of the dataset is normalized using its standard deviation and fed to the clustering algorithm built-in the Sklearn package [63]. Data vectors used for clustering are selected by applying weights ranging between 0 and 1. For example, clustering can be performed by selecting the position of the point (x, y) and its friction by applying a zero weight to the topography data, eliminating the contribution of this part of the dataset to the definition of the clusters. The clustering algorithm employs Radial Basis Function (RBF) Gaussian kernels to build the affinity matrix of the dataset [21, 58]. The principal components of the unnormalized Laplacian of the affinity matrix are obtained by numerically computing the eigenvalues and associated eigenvectors. K-means or a discretization algorithm based on Single Value Decomposition (SVD) are applied on the decomposition along principal components to build the different clusters.

In our implementation, 2 or 3 clusters are normally employed. While one clustering step is sufficient to segment fully most of our datasets, a recursive method can be employed when the signal to noise ratio is particularly low or when different clusters have very different distributions, we refer to the Supplementary Information for further details on the procedure.

**Å-indentation for transverse elasticity of 2D layers**

Å-indentation is based on Modulated NanoIndentation (MoNI) [10, 22, 23] AFM and allows sub-Ångström resolution indentation measurements. This technique is particularly effective in measuring elastic properties of 2D materials [22], thanks to indentation depths that are comparable to or smaller than the distance between the 2D layers (sub- Å). Å-indentation experiments have enabled direct measurement of the interlayer properties of epitaxial graphene and graphene oxide [22], as well as the discovery of the ultra-stiff diamene film on few-layers epitaxial graphene [10]. For a comprehensive analysis of this technique we refer to the literature [64]. For the here presented Å-indentation results, we used a sinusoidal voltage (<0.4 mV) applied using a lock-in amplifier (Stanford Research Systems, SR830) to the piezotube of the AFM (Agilent PicoPlus AFM) to drive small oscillations ($\Delta z_{piezo}$~0.1 Å) in parallel to the main AFM driving voltage. Å-indentation experiments are conducted on epitaxial graphene on SiC as well as on graphene flakes on silicon oxide using Nanosensors DT-NCHR tips (spring constant ~80 N/m, tip radius ~100-200 nm). Pressures applied by the tip on the surface of the sample during the Å-indentation experiments are estimated to be in the range 3-7 GPa, which is comparable to values reported in Reference [10]. The indentation modulus measured for the 2D films is compared with the stiffness of the substrate as well as the stiffness of other ultra-stiff reference materials, namely CVD (001) diamond film obtained through a HFCVD process [65], SiC (0001), and bulk (0001) sapphire [11].

## Acknowledgments

The Authors acknowledge the support from the Office of Basic Energy Sciences of the US Department of Energy (grant no. DE-SC0018924).

## Author contributions

F.C. prepared exfoliated graphene samples, conducted nanomechanics experiments on exfoliated graphene, developed the spectral clustering algorithm, and performed data analysis. F.L. conducted nanomechanics experiments on epitaxial graphene and analyzed the data. E.R. conceived and designed the experiments and analyzed the data. C.B and W. de H. prepared the epitaxial graphene samples. All the Authors contributed to writing the manuscript.

24. Ferrari, A.C., et al., *Raman spectrum of graphene and graphene layers.* Physical Review Letters, 2006. **97**(18).
25. Ferrari, A.C. and D.M. Basko, *Raman spectroscopy as a versatile tool for studying the properties of graphene.* Nature Nanotechnology, 2013. **8**(4): p. 235-246.
26. Li, H., et al., *Rapid and reliable thickness identification of two-dimensional nanosheets using optical microscopy.* ACS Nano, 2013. **7**(11): p. 10344-10353.
27. Blake, P., et al., *Making graphene visible.* Applied Physics Letters, 2007. **91**(6).
28. Rejhon, M. and J. Kunc, *ZO phonon of a buffer layer and Raman mapping of hydrogenated buffer on SiC (0001).* Journal of Raman Spectroscopy.
29. Rohrl, J., et al., *Raman spectra of epitaxial graphene on SiC(0001).* Applied Physics Letters, 2008. **92**(20).
30. Robinson, J.A., et al., *Correlating Raman spectral signatures with carrier mobility in epitaxial graphene: a guide to achieving high mobility on the wafer scale.* Nano letters, 2009. **9**(8): p. 2873-2876.
31. Röhrl, J., et al., *Raman spectra of epitaxial graphene on SiC (0001).* Applied Physics Letters, 2008. **92**(20): p. 201918.
32. Lee, C., et al., *Measurement of the elastic properties and intrinsic strength of monolayer graphene.* Science, 2008. **321**(5887): p. 385-388.
33. Lee, C., et al., *Frictional characteristics of atomically thin sheets.* Science, 2010. **328**(5974): p. 76-80.
34. Lee, H., et al., *Comparison of frictional forces on graphene and graphite.* Nanotechnology, 2009. **20**(32).
35. Paolicelli, G., et al., *Nanoscale frictional behavior of graphene on SiO2 and Ni(111) substrates.* Nanotechnology, 2015. **26**(5).
36. Berger, C., et al., *Electronic confinement and coherence in patterned epitaxial graphene.* Science, 2006. **312**(5777): p. 1191-1196.
37. Kruskopf, M., et al., *Comeback of epitaxial graphene for electronics: large-area growth of bilayer-free graphene on SiC.* 2D Materials, 2016. **3**(4).
38. Forti, S. and U. Starke, *Epitaxial graphene on SiC: from carrier density engineering to quasi-free standing graphene by atomic intercalation.* Journal of Physics D: Applied Physics, 2014. **47** p. 094013.
18

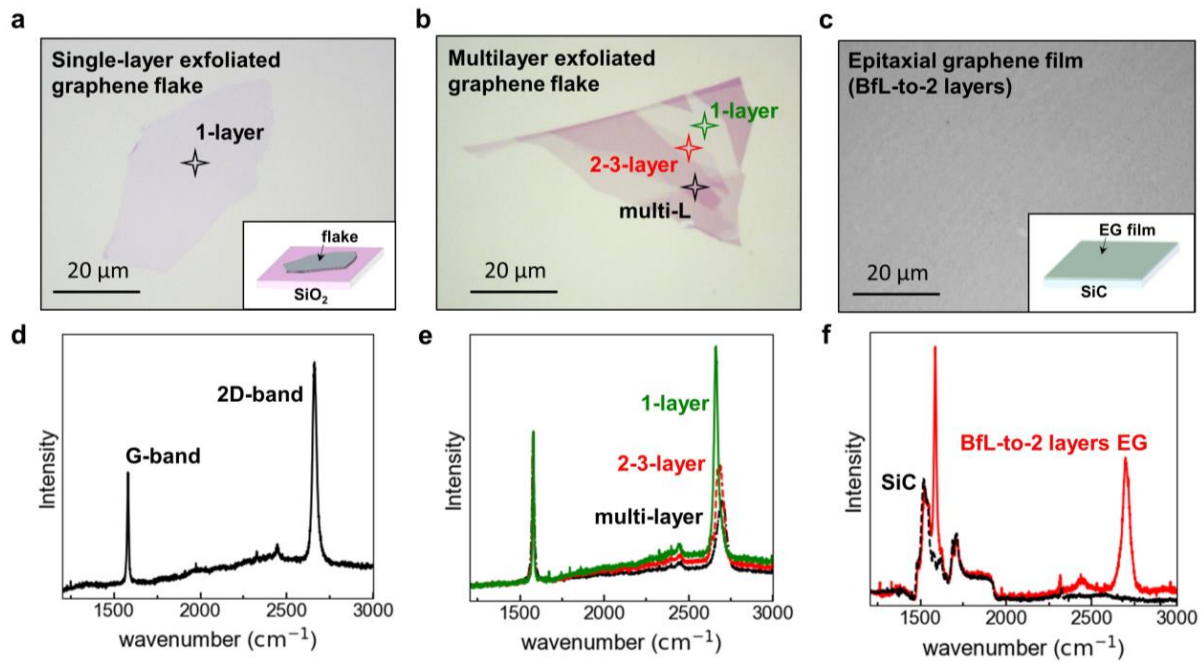

**Fig. 1. Optical micrographs and Raman spectra of exfoliated and epitaxial graphene.** (a) Optical micrograph of exfoliated graphene single-layer (1-layer) flake on silicon oxide. (b) Optical micrograph of exfoliated graphene multilayer flake on silicon oxide. (c) Optical micrograph of epitaxial graphene on silicon carbide. (d) Raman spectrum of 1-layer graphene. (e) Raman spectrum of multilayer graphene flake at different locations including single-layer (1-layer), few layer graphene (2-3-layer), and multilayer (multi-layer) graphene. Markers in (b) indicate the approximate location where spectra are measured. (f) Raman spectrum of epitaxial graphene film (BfL-to-2 layers) on Si-face of silicon carbide and bare silicon carbide substrate.

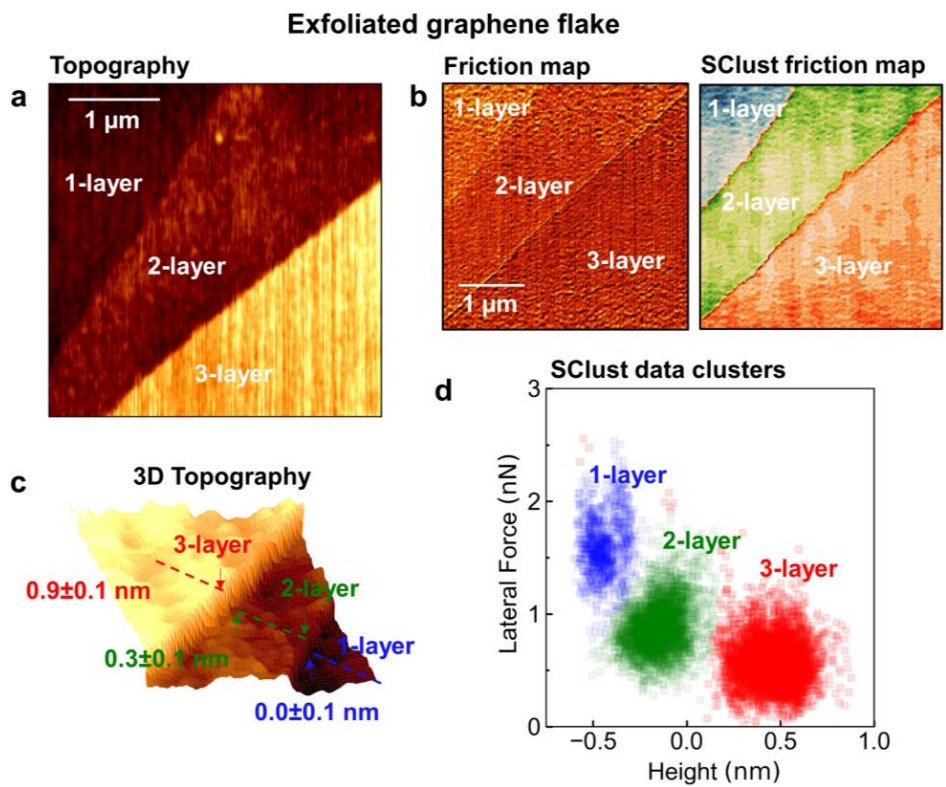

**Fig. 2. Spectral clustering of topography and FFM data of exfoliated graphene.** (a) Topography for 1-, 2-, 3-layer exfoliated graphene film. (b) Friction (FFM) map of exfoliated 1-, 2-, 3-layer graphene and machine learning (SClust) friction map obtained by processing the scan data through the spectral clustering algorithm described in Methods. Shaded areas corresponds to 1-layer (blue), 2-layer (green), and 3-layer (red) graphene, as determined through spectral clustering. (c) 3D topography of the exfoliated film. Numerical values reported on the topography are average heights of the layers with respect to the mid plane of the 1-layer domain. Error margins are one standard deviation from the mean. (d) Lateral force and topographic height for the 1-layer, 2-layer, and 3-layer graphene domains extracted from the topography (a) and friction force (b) maps using the clustering algorithm. Distributions are obtained by processing the scans through the spectral clustering algorithm, as described in the Methods section.

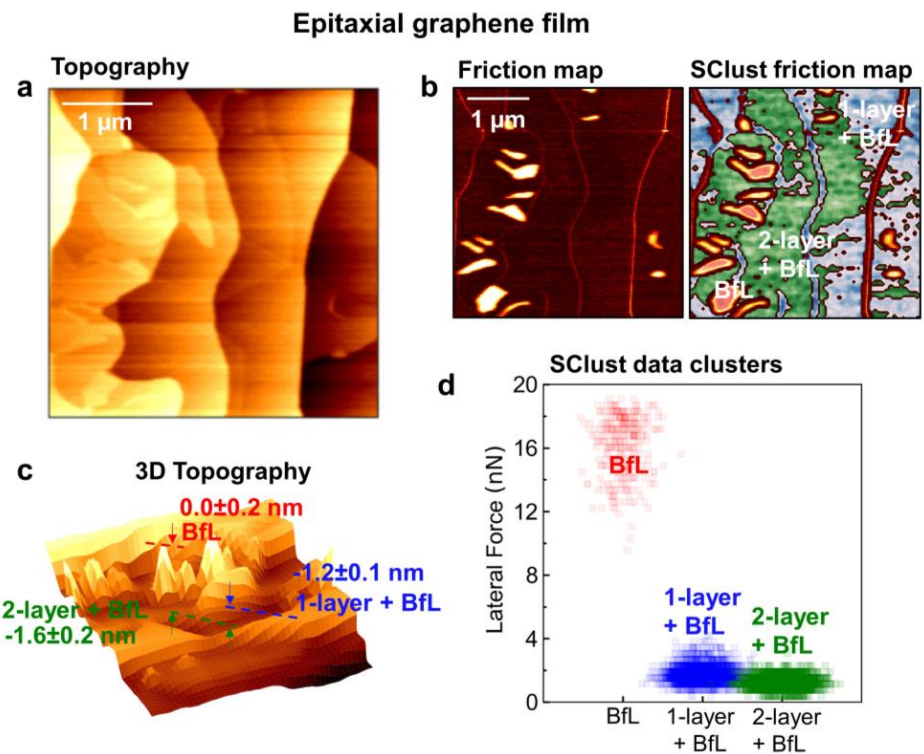

**Fig. 3. Spectral clustering of topography and FFM data of epitaxial graphene films.** (a) Topography of the epitaxial graphene film. (b) Friction (FFM) map of epitaxial graphene film and machine learning (SClust) friction map obtained by processing the scan data through the spectral clustering algorithm described in Methods. Shaded areas corresponds to 1-layer graphene (blue), 2-layer graphene (green), and BfL (red), as determined through spectral clustering. (c) 3D topography of the epitaxial graphene film. Numerical values reported on the topography are average heights of the BfL, 1-layer and 2-layer graphene regions with respect to the mid plane of the BfL domain. Error margins are one standard deviation from the mean. (d) Lateral force for the 1-layer graphene, 2-layer graphene, and BfL domains extracted from friction force (b) maps using the clustering algorithm. Distributions are obtained by processing the scans through the spectral clustering algorithm as described in Methods.

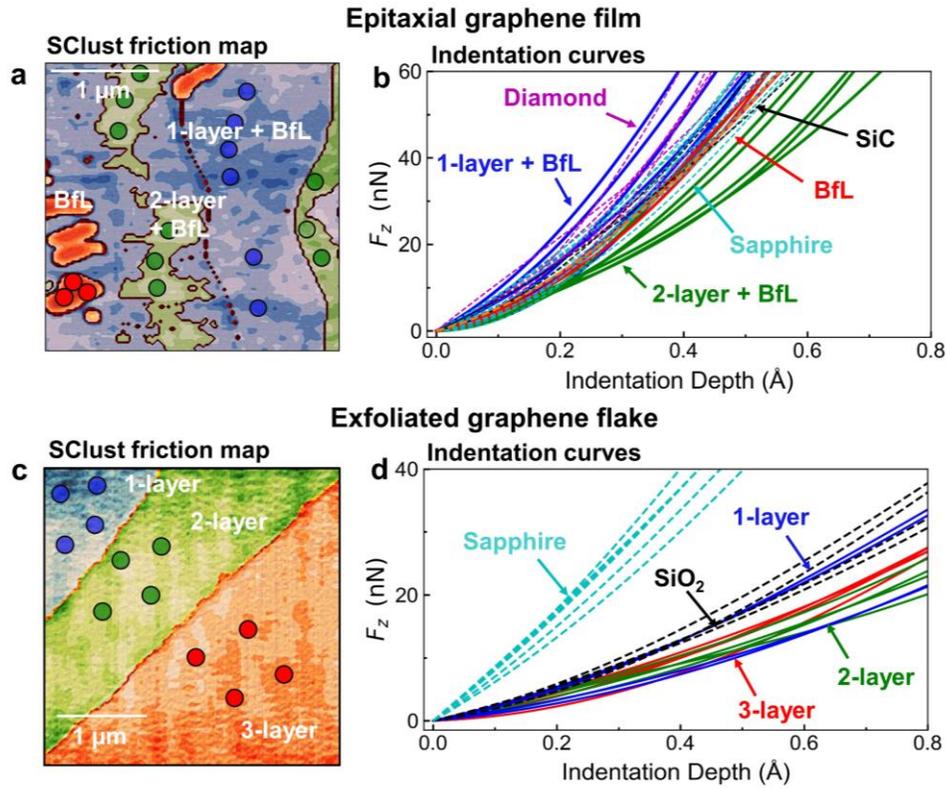

**Fig. 4. Å-indentation of few-layer epitaxial and exfoliated graphene.** (a) SClust friction map of epitaxial graphene surface. Markers indicate the positions where Å-indentation experiments are conducted. (b) Indentation curves obtained in the positions indicated in (a) for the 1-layer (blue), 2-layer (green), and BfL (red) domains identified in the scan (a). For direct comparison, indentation curves measured on reference materials CVD diamond, SiC, and sapphire are also reported. (c) SClust friction map of exfoliated graphene flake 1-, 2-, 3-layer. Markers indicate the positions where Å-indentation experiments are conducted. (d) Indentation curves obtained in the positions indicated in (a) for the 1-layer (blue), 2-layer (green), and 3-layer (red) domains identified in the scan (a). Indentation curves measured on $SiO_2$ in a different region of the same sample are also reported as well as indentation curves measured on sapphire, which is used as an external reference.

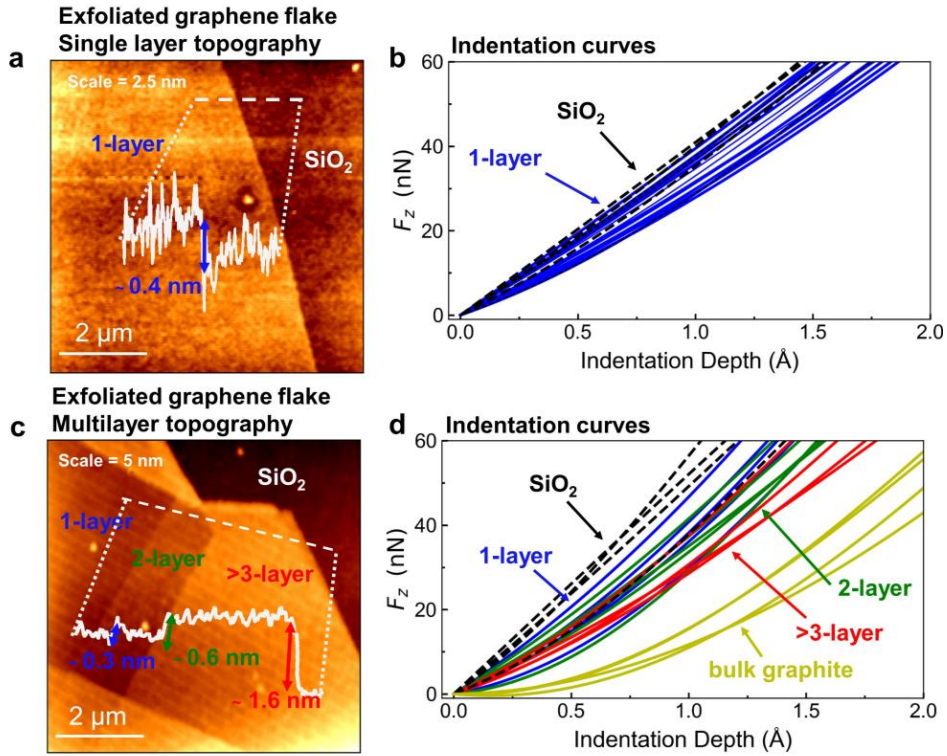

**Fig. 5. Å-indentation of single layer and multi-layer exfoliated graphene.** (a) Topographic scan of 1-layer exfoliated graphene on $SiO_2$. In the graphic, surface topography along the dashed line showing the height of the 1-layer flake is displayed. (b) Å-Indentation curves measured on $SiO_2$ and 1-layer graphene. (c) Topographic scan of graphene multi-layer flake on $SiO_2$. In the graphic, graphene topography along the dashed line showing the height difference between the different regions of the flake. (d) Indentation curves measured on $SiO_2$ and different regions of the graphene flake, namely 1-layer (blue), 2-layer (green), and >3-layer (red). Indentation curves measured on bulk graphite (>>10 layer graphene, yellow) and bare $SiO_2$ (black) are also reported.

# Supplementary Information: Layer dependence of graphene-diamene phase transition in epitaxial and exfoliated few-layer graphene using machine learning

Filippo Cellini[1§], Francesco Lavini[1,2§], Claire Berger[3,4], Walt de Heer[4,5], and Elisa Riedo[1*]

[1] Tandon School of Engineering, New York University, Brooklyn, NY, USA.

[2] Graduate School of Art and Sciences, New York University, New York, NY, USA.

[3] Institut Néel, CNRS-University Grenoble-Alpes, Grenoble, France

[4] School of Physics, Georgia Institute of Technology, Atlanta, GA, USA

[5] TICNN, Tianjin University, Tianjin, China

[*] Corresponding author: elisa.riedo@nyu.edu

§ These Authors contributed equally (see Authors contributions)

## S1. Spectral clustering of exfoliated and epitaxial graphene phase imaging data

The spectral clustering algorithm is employed to analyze phase imaging data acquired during tapping mode experiments. Diamond coated tips are employed in phase imaging experiments with resonant frequencies in the range 400-500 kHz. In order to control the interaction of the tip with the sample, the amplitude of the oscillation is modulated in the range 0.4-0.6 $A_0$, with $A_0$ being the amplitude of the driving oscillation in air (amplitude of oscillation is set to 6-8 V on our Agilent Picoplus AFM). Due to the presence of a variable offset in the phase measurement, phase values are reported as phase shifts between the graphene single layer (1-layer epitaxial and exfoliated graphene) and the other graphene structures (2-3-layer) and buffer layers.

The spectral clustering algorithm discussed in the Methods section is used to isolate graphene domains in phase imaging data coupled with topographic scans in both exfoliated graphene flakes and epitaxial graphene films. Phase shift data for the exfoliated graphene flake are displayed in Figure S1(a) (topography in Figure 2 in the paper), together with the clustered regions identified through the spectral clustering algorithm. Notably, separation between the different regions is less defined than in the case of FFM data in Figure 2 (in the paper), whereby a smooth increase is observed with increasing number of layers from 1-layer to 3-layer. In addition, experiments conducted on $SiO_2$ and 1-layer and 3-layer graphene (not reported) show that a small variation is observed in the phase shift between $SiO_2$ and 1-layer graphene [1], while a measurable increase is observed between 1-layer and 3-layer. Quantitative evaluation of the phase shift is conducted from clustered data, with the first domain corresponding to 1-layer graphene presenting an average phase shift of 0 (1-layer is assumed as the reference as discussed above) with an associated standard deviation of 1.9 deg. The relative height of the region, which is assumed as the reference plane in the scan, is 0.0 nm with standard deviation of 0.2 nm. The 2-layer region has average phase shift of 3.0-±2.3 deg and average height of 0.8±0.3 nm. The 3-layer graphene region has average phase shift of 4.1±1.8 deg and average height of 1.4±0.3 nm.

Spectral clustering algorithm is also applied to phase shift data for epitaxial graphene films. Phase data are displayed in Figure S1(b) together with the clustered regions (topography in Figure 3 in the paper). Phase shift data for the three domains are reported in Figure S1(d). The region corresponding to the buffer layer (BfL) presents an average phase shift of 8.7 deg with an associated standard deviation of 1.5 deg. The region corresponding to 1-layer graphene presents an average phase shift of 0 in the graph, as 1-layer graphene is again used as the reference in phase imaging experiment, with a standard deviation of 0.8 deg. The 2-layer graphene region has average phase shift of 3.3±1.2 deg.

## S1. Recursive spectral clustering for graphene AFM data

In the case of noisy datasets or datasets where clusters have particularly different distributions, a recursive approach is selected to identify the graphene domains. In this method, 2 clusters are initially identified through the clustering algorithm. The cluster with the largest number of elements is retained for successive spectral clustering steps, and the affinity matrix is computed on a reduced dataset composed only of data points of the largest cluster. The procedure is repeated iteratively. At each step the variation of the gap between the eigenvalues (eigengap) is evaluated to empirically verify that the "quality" of the clustering is increasing, and the procedure is considered completed when a marginal increase in eigengap is obtained between two steps.

Notably, the distance between eigenvalues can be regarded as a metric to evaluate the quality of the spectral clustering results [2]. This procedure, based on a simple decision tree method, is largely empirical, but it is shown to give good results on our dataset in cases in which the direct application of the spectral clustering algorithm with more than 2 clusters does not define reasonable graphene domains. In Figure S2, we display the recursive clustering procedure applied to one of the FFM experiments discussed in the paper. Figure S2(a) displays the friction map measured for an epitaxial graphene film. The data from the scan are fed to the recursive clustering algorithm and at each step 2 clustered regions and the associated eigenvalues of the Laplacian of the affinity matrix are computed. The distance between the eigenvalues (eigengap) computed at each clustering step is displayed in Figure S2(b). The clustered regions computed at each step are displayed in Figure S2(c). The clustering obtained at step 3 is identified as the optimal clustering for this experiment, whereby the increase in the eigengap between step 3 and step 4 is small, and therefore the relative improvement in the quality of the spectral clustering is considered to be small. Clustered regions at step 3 show a satisfactorily distinction between the three graphene regions, namely BfL, 1-layer, and 2-layer epitaxial graphene. This recursive procedure, while defined empirically for our experiments, is proved to work effectively on our datasets.

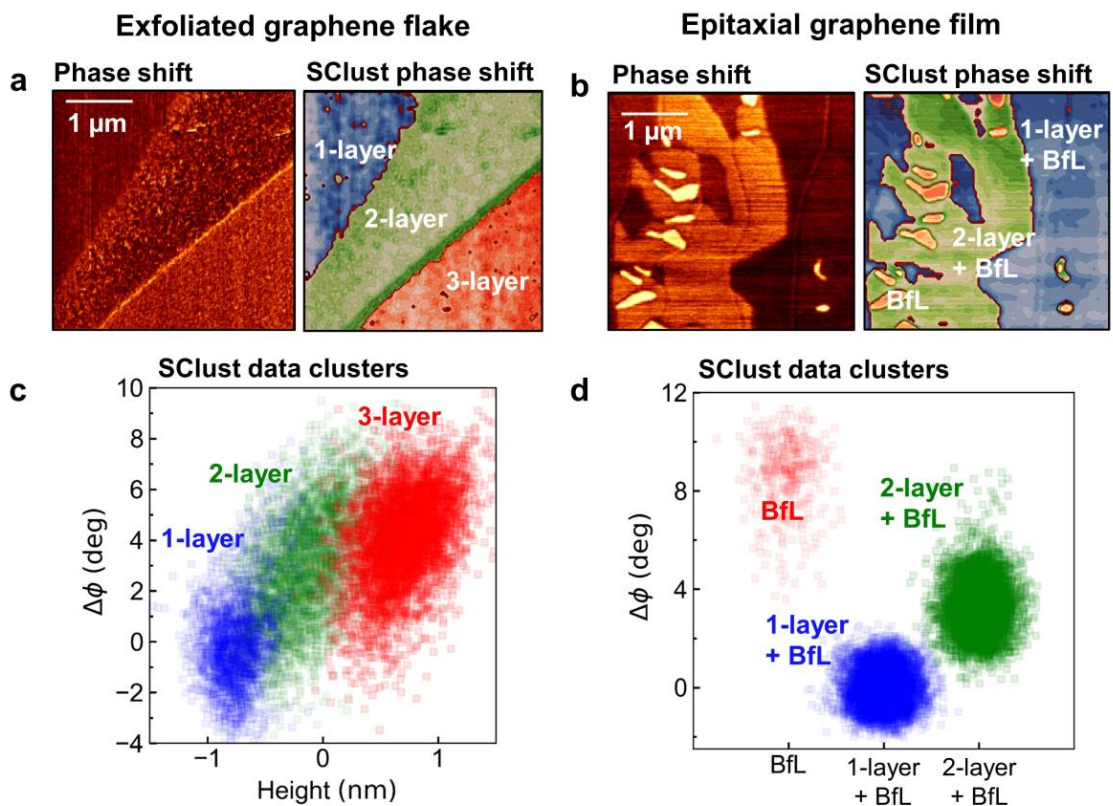

**Fig. S1. Spectral clustering of Topography and phase shift data of exfoliated graphene flake and epitaxial graphene film.** (a) Phase shift map of exfoliated 1-, 2-, 3-layer graphene and machine learning (SClust) friction map obtained by processing the scan data through the spectral clustering algorithm described in Methods. Shaded areas corresponds to 1-layer (blue), 2-layer (green), and 3-layer (red) graphene, as determined through spectral clustering. (b) Phase shift and topographic height for the 1-layer, 2-layer, and 3-layer graphene domains extracted from the topography (Figure 2) and phase shift (a) maps using the clustering algorithm. Distributions are obtained by processing the scans through the spectral clustering algorithm, as described in Methods. (b) Phase shift map of epitaxial graphene film and machine learning (SClust) friction map. Shaded areas corresponds to 1-layer graphene (blue), 2-layer graphene (green), and BfL (red), as determined through spectral clustering. (d) Lateral force for the 1-layer graphene, 2-layer graphene, and BfL domains extracted from phase shift map (b) using the clustering algorithm.

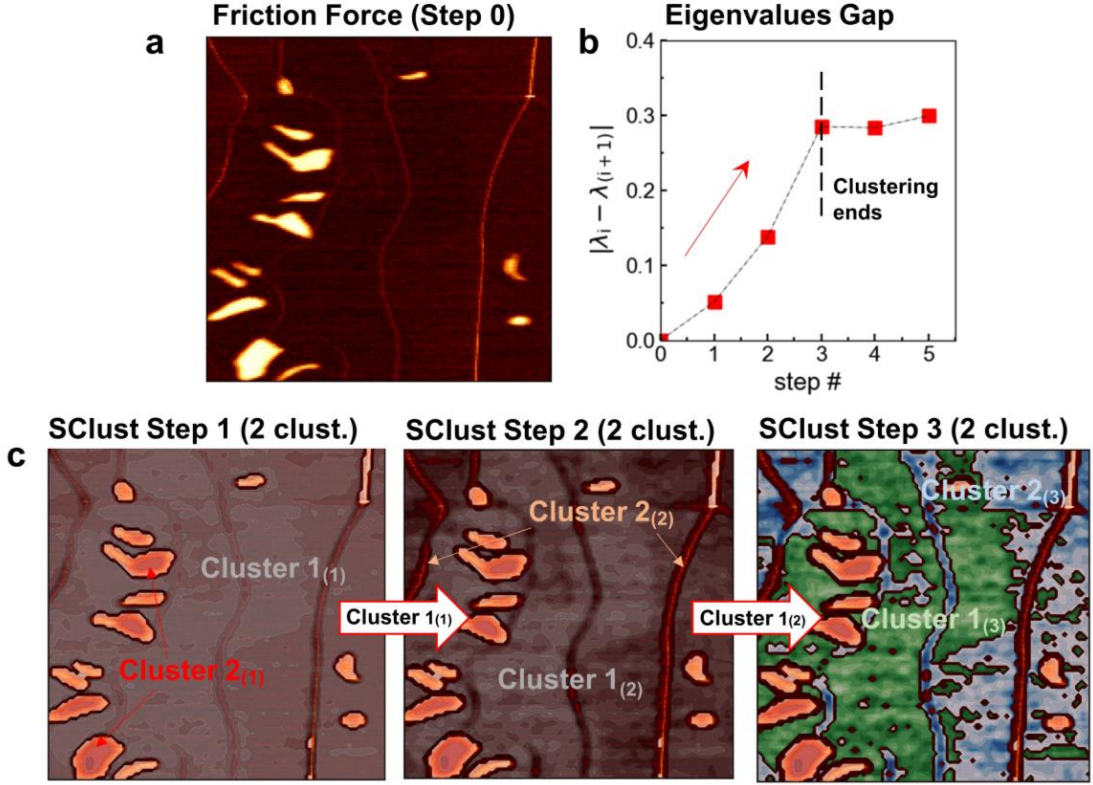

**Fig. S2. Recursive spectral clustering of topography and FFM data of epitaxial graphene film.** (a) (a) Friction map of epitaxial graphene film, please see Figure 3 in the paper and the Results and Methods section for details on the FFM measurements. (b) Difference between the two eigenvalues determined at each step for the unnormalized Laplacian of the affinity matrix using the Eigsh function in the Scipy package in Python. The difference between the eigenvalues is empirically used as a metric of the quality of the spectral clustering. (c) Clustered regions at steps 1 to 3 in the recursive spectral clustering algorithm. At each step, the spectral clustering is performed on the largest cluster between the two clusters identified in the previous step. The cluster are identified using the notation Cluster N(m) where N is the Cluster number and m is the Step number. The optimal clustering is identified at step 3 as indicated in (b).